\documentclass[12pt]{JHEP3}
\usepackage{amsmath,amssymb}

\title{Boundary conditions as constraints}

\author{Juan M. Romero and J. David Vergara \\
Instituto de Ciencias Nucleares, U.N.A.M., A. Postal 70-543,
M\'exico D.F., M\'exico
E-mail: \email{sanpedro@nuclecu.unam.mx}, \email{vergara@nuclecu.unam.mx} }

\abstract{
A new method to compute the symplectic structure of a quantum
field theory with non trivial boundary conditions is proposed.
Following the suggestion in \cite{ho:gnus, ardalan}, we regard
that the boundary conditions are second class constraints in the
sense of the Dirac's method. However, we show that this proposal
is more useful if we consider an inverse of the Holographic map
between a theory defined in the boundary to another with
constraints but without boundary.}

\begin{document}

\newcommand{\rof}[1]{(\ref{eq:#1})}

\section{Introduction}
In recent years, the study of the boundary conditions in Quantum
Theory has produced several important results. For example, an
interesting idea is to consider the black hole event horizon as a
physical boundary \cite{car:gnus}, this induce an extra term in
the action, having as consequence the existence of a central
charge in the algebra of generators of gauge transformations
\cite{br:gnus}. Using this central charge it is possible to
determine the asymptotic behavior of the density of states and in
this way to get the entropy for a black hole in 2+1-dimensions
\cite{car1:gnus}.

In the context of string theory, the D-branes  are natural
boundaries for the open strings. These boundaries have very
interesting effects in the theory among them is the
non-commutativity of the D-brane coordinates \cite{W:gnus,
Sz:gnus}. This non commutativity appears due to the presence of a
constant B-field on the boundary that implies a change of the
symplectic structure on the boundary. This change of the
symplectic structure has been computed using different methods,
the most simple is the direct solution of the Field equations
subject to the Dirichlet boundary conditions on the 9-p transverse
directions and Neumann boundary conditions on p+1 directions
parallel to the Dp-brane. However, for some interesting systems is
not possible to solve exactly this problem and the question
appears if is there exist other method to compute the symplectic
structure.

An alternative to solve this problem was to consider that the
boundary conditions in a Quantum Field Theory can be interpreted
as second class constraints  in the sense of the Dirac's Method
\cite{ho:gnus, ardalan}. This procedure has some interesting
characteristics, the non commutativity appears naturally due to a
modification of the symplectic structure given by the Dirac's
brackets. These brackets are constructed with the second class
constraints that arise from the boundary conditions and the time
evolution of these conditions. However, the procedure have some
problems, see for example \cite{Loran:gnus}.  First the Lagrange
multipliers are fixed by hand and not following the standard
Dirac's method where the Lagrange multipliers are fixed in the
case of second class constraints by the time evolution of the
constraints. This has as consequence that appears an infinite
number of constraints and then a minus infinite number of degrees
of freedom in the boundary. To solve these problems we propose an
alternative procedure that follows in many aspects the previous
proposal. However we have a very different interpretation that
allows solve most of the shortcomings. The key point of our
procedure is to perform a mapping from the original problem with a
given Hamiltonian and boundary conditions to another problems with
the same Hamiltonian, but now with second class constraints equal
to the boundary conditions  and no boundary. Our conjecture is
that the results that we get for the Dirac's brackets projected to
the interior of the new problem, is the symplectic structure of
the original problem. We show that this conjecture is valid in all
examples that we know and inclusive we check that in the example
of PP-waves \cite{ho2:gnus} our result is fully consistent whereas
the previous result it is not. In the literature exist several
proposals \cite{jab:gnus, san2:gnus, san3:gnus, san4:gnus,
san5:gnus, san6:gnus, san7:gnus, He:gnus, Banerjee:gnus,
Deriglazov:gnus} to solve the problems associated to the procedure
of consider the boundary conditions as Dirac's constraints.
However, all the previous proposals are not useful  in at least in
one of the examples that we present, whereas our procedure is
fully consistent and systematic  in all these examples.

In section 2 we present an outline of our procedure, in the next
section we analyze the case of the scalar field with Neumann and
Dirichlet boundary conditions. In section 4 we present the case of
the bosonic string with mixed boundary conditions and in section 5
we study the case of the bosonic string in a PP-wave background.

\section{Boundary conditions as constraints}

\subsection{Boundary conditions and the action principle}
\label{2}

Let $M$ be a $(d+1)$-dimensional manifold, with topology $\Sigma
\times \mathbb{R}$, where $\Sigma$ is an oriented $d$-manifold
with boundary $\partial \Sigma $. In $M$ we consider a Field
Theory given by the action
\begin{eqnarray}\label{A1}
S=\int_{M} dx^{d+1} {\it L}(\phi_{a}(x),
\partial_{\alpha}\phi_{a}(x)),
\end{eqnarray}
for the fields $\phi_a$. The conditions that the integral
(\ref{A1}) be stationary implies
\begin{equation}\label{varS}
\begin{split}
 \delta S= &\int_{M} dx^{d+1} [\frac{\partial {\it L}}{\partial
\phi_{a}(x)}-\partial_{\alpha}(\frac{\partial {\it
L}}{\partial(\partial \partial_{\alpha}\phi_{a}(x))})]\delta
\phi_{a}(x) \\
& \ +\int_{M} dx^{d+1}
\partial_{\alpha}[(\frac{\partial {\it L}}{\partial(
\partial_{\alpha}\phi_{a}(x))})\delta \phi_{a}(x)] =0.
\end{split}
\end{equation}
For a system without boundary the second term cancel
automatically, but in our case we need to impose boundary
conditions. There are three different ways to do that:

Dirichlet conditions:

\begin{equation}\label{diri}
\delta \phi_{a}(x)|_{x\epsilon \partial\Sigma}=0
\end{equation}

Neumann conditions:

\begin{equation}\label{neum}
\left(\frac{\partial {\it L}}{\partial(
\partial_{\alpha}\phi_{a}(x))}\right)(x)|_{x\epsilon
\partial\Sigma}=0,
\end{equation}
or one combination of both types for the components of the field
$\phi_a$.

\vskip1pc

On the other hand, we have  the canonical formalism defined by the
Hamiltonian,

\begin{eqnarray}
H_{c}=\int_{\Sigma}dx^{d}{{\cal H}_{c}}(\phi, \partial \phi, \Pi,
\partial \Pi),\label{eq:cano}
\end{eqnarray}
the symplectic structure,
\begin{eqnarray}
\{\phi_{a}(x,t),\phi_{b}(x^{\prime},t)\}=\{\Pi_{a}(x,t),\Pi_{b}
(x^{\prime},t)\}=0\label{eq:lechuga0}\\
 \{\phi^{a}(x,t),\Pi_{b}(x^{\prime},t)\}=
\delta^{a}_{b}\delta(x-x^{\prime})\label{eq:lechuga01},
\end{eqnarray}
and the boundary conditions,
\begin{equation}\label{BC1}
F_{a}(\phi, \partial \phi, \Pi, \partial \Pi)|_{x\epsilon \partial
\Sigma}=0.
\end{equation}
This set of boundary conditions is equivalent in the canonical
formalism to the equations (\ref{diri}) or (\ref{neum}). In the
definition of the canonical formalism we are assuming that the
boundary conditions are not in contradiction with the local
symmetries of the system. So from the beginning we consider that
our system does not have gauge degrees of freedom or that a gauge
choice consistent with the boundary conditions has been done.
Also, we are assuming that the symplectic structure
(\ref{eq:lechuga0})-(\ref{eq:lechuga01}) is defined until a  zero
measure set, so this structure can be changed in the boundary
without affect the rest of the theory. So, if the boundary
conditions (\ref{BC1}) are not consistent with the symplectic
structure, it is possible to modify the structure in the boundary
in such way that be consistent with the boundary conditions. For
example suppose that we have the boundary conditions,
$$F_{a}=[\Pi_{a}(x)-\phi_{a}(x)]|_{x\epsilon \partial\Sigma}=0,$$
these conditions are clearly inconsistent with
(\ref{eq:lechuga01}), but we can modify the structure in the
boundary, so we can introduce
\begin{equation}\label{SSM1}
  \{\phi^{a}(x,t),\Pi_{b}(x^{\prime},t)\}|_{x\epsilon \partial\Sigma}=
  0.
\end{equation}
This symplectic structure is consistent with the boundary
conditions. In principle, there are many ways to modify the
symplectic structure in such way that these be consistent with the
boundary conditions. However, no all of these modified structures
are consistent with the exact solution of the equations of motion
of the system. The main result of our paper is to propose a new
procedure to obtain a symplectic structure that agrees with the
symplectic structure obtained by the exact solution of the
problem.

\subsection{Outline of the method to compute the symplectic
structure in the boundary}

Following the reference  \cite{ho:gnus, ardalan, jab:gnus}, we
assume that the boundary conditions are primary constraints in the
sense of the Dirac's method. However, we want to construct a
procedure fully consistent with the Dirac's method, avoiding
problems with regularization. To achieve this we consider that the
primary constraints are not only valid in the boundary
$\partial\Sigma$. Then, we suppose that these constraints are
valid on all $\Sigma$, i.e.,

\begin{equation}\label{PC}
    F_{a}(\phi, \partial \phi, \Pi, \partial \Pi)
|_{x \epsilon \Sigma}\approx 0.
\end{equation}
So, it make sense to write these constraints in smeared form
\begin{equation}\label{DPC}
   F[{\bf N}]=\int_{\Sigma} dx^{d}N^a(x)F_{a}(\phi,
\partial \phi, \Pi, \partial \Pi)\approx 0,
\end{equation}
for $N^a(x)$ smeared functions with compact support. In
consequence, we are mapping our original problem with boundary
conditions (\ref{BC1}) and canonical Hamiltonian (\ref{eq:cano})
to another problem where we have primary constraints (\ref{PC}) an
a total Hamiltonian, given by
\begin{equation}\label{TH}
    H_{T}=H_{c}+\int_{\Sigma}dx^{d}\lambda^{a}F_{a}.
\end{equation}
Using the total Hamiltonian (\ref{TH}) we evaluate the time
evolution of the primary constraints,
\begin{equation}\label{EPC}
    \dot {F}[{\bf N}] =[F[{\bf N}], H_T].
\end{equation}
From this procedure we can obtain new constraints. To this
secondary constraints, we applied the same procedure, we evolve
the constraints and see if we get new constraints. We finish the
procedure when we don't get new constraints and determine the
Lagrange's multipliers. This must be so, because we are
considering that in the original problem (\ref{A1}) does not have
gauge degrees of freedom and furthermore that the boundary
condition (or primary constraints), do not generate gauge freedom
in the boundary. We conclude that the system only have second
class constraints, then all Lagrange's multipliers associated to
the primary constraints must be determined. The set of second
class constraints that we obtain are denote by,
\begin{equation}\label{SCC}
    \chi_{\alpha}(\phi,\partial\phi, \pi, \partial\pi)|_{x\in
    \Sigma} \approx 0.
\end{equation}
With these constraints we obtain the invertible matrix
\begin{equation}\label{IC}
    C_{\alpha \beta}(x,x^{\prime})=
    \{\chi_{\alpha}(x),\chi_{\beta}(x^{\prime})\}
\end{equation}
and using (\ref{IC}) we construct the Dirac bracket,
\begin{equation}\label{DB}
\{A(x),B(x^{\prime})\}^{*}=\{A(x),B(x^{\prime})\}-\{A(x),
\chi_{\alpha}(y)\}C^{\alpha \beta}(y,z)\{\chi_{\alpha}(z),
B(x^{\prime})\}.
\end{equation}
We notice that the Dirac bracket (\ref{DB}) is valuated on the
surface $\Sigma$, and it is not smeared. So, to obtain a Dirac
bracket that is valid only in the interior of sigma $\Sigma -
\partial\Sigma$, we need to eliminate from the computation of the
bracket the part that is different of zero only in the boundary
$\partial\Sigma$. Finally, we affirm that the correct symplectic
structure in the boundary is obtained from the projection of this
bracket to the boundary. Our procedure is in some sense equivalent
to a Wick rotation, where one maps the original problem to the
Euclidean space to make sense to the integrals, here we map the
original problem defined with a boundary to another problem
without boundary but with primary constraints that reflects the
presence of the boundary. We do that in order to make sense of the
Poisson brackets and in this way avoiding the problems with the
regularization and obtaining a Dirac's method fully consistent.
Finally, when we obtain the symplectic structure in the new
problem it is possible to obtain the symplectic structure of the
original problem by projecting the structure resulting from the
Dirac brackets to the boundary. We see, that our procedure is in
some sense equivalent to an inverse of the Holographic map, since
we obtain information about the boundary using computations in the
bulk.

We don't have a rigorous proof that the outlined procedure is
always correct. However, we show in the following section that for
several examples that our results coincide with the exact results
obtained using the exact solution of the problem. In the example
of the bosonic string in a background of PP-waves we disagree with
one part of the result recently published. Nevertheless, we show
that our symplectic structure is fully in the boundary whereas the
result of \cite{ho2:gnus} is inconsistent in this region.

\section{The scalar field with Dirichlet and Neumann conditions} We
introduce our procedure in a very simple example, the scalar
field. This example is quite useful because we have an exact
solution for boundary conditions of the type of Dirichlet and
Neumann, and we show that our method reproduce correctly the
symplectic structure in both cases. We start from the action

\begin{eqnarray}\label{ASF}
S=\frac{1}{2}\int_{0}^{\pi}\int_{t_{1}}^{t_{2}}dxdt[ (\partial_{t}
\phi(x,t))^{2}-(\partial_{x} \phi(x,t))^{2}].
\end{eqnarray}
From this action we get the equations of motion
\begin{equation}\label{EMSF}
    (\partial^{2}_{t}- \partial^{2}_{x})\phi(x,t)=0,
\end{equation}
and we have two possible choices of boundary conditions
\begin{eqnarray}
\phi_{i}(x,t)|_{x=\pi, 0}&=0 \qquad \quad &{\rm Dirichlet},\label{DSF} \\
(\partial_{x}\phi_{i})(x,t)|_{x=\pi, 0}&=0 & {\rm
Neumann}\label{NSF}.
\end{eqnarray}
and the canonical Hamiltonian is given by
\begin{eqnarray}
H_{c}=\frac{1}{2}\int_{0}^{l}dx[(\Pi(x,t))^{2}
+(\partial_{x}\phi(x,t))^{2}].\label{eq:jitomate}
\end{eqnarray}

\subsection{ Scalar Field with Dirichlet boundary conditions}
For the Dirichlet boundary condition we have the primary
constraint
\begin{equation}\label{PCDSF}
    \phi(x,t)\approx 0.
\end{equation}
That we can rewrite in densityzed form as,
\begin{equation}\label{DPCDSF}
    \Phi_D^{(1)}(N)=\int_{0}^{\pi}dxN(x)\phi(x,t)\approx 0,
\end{equation}
and the total Hamiltonian is,
\begin{equation}\label{THSFD}
    H_{T}=H_{c}+\int_{0}^{\pi}dx\lambda(x)\phi(x,t).
\end{equation}
From, the time evolution of the primary constraint (\ref{PCDSF}),
we get
\begin{equation}\label{TEDSF}
    \dot {\Phi}_D^{(1)}(N) = \{ \Phi^{(1)}(N), H_T \} = \int_0^\pi
    dx  N(x) \pi(x) \approx 0 ,
\end{equation}
which implies that we have the secondary constraint,
\begin{equation}\label{SCDSF}
     \Phi_D^{(2)}(M)=\int_{0}^{\pi}dxM(x)\Pi(x,t)\approx 0.
\end{equation}
The time evolution of the constraint (\ref{SCDSF}) produce no new
constraints and we get the Lagrange multiplier,
\begin{equation}\label{LDSF}
    \lambda=\partial^{2}_{x}\phi(x,t).
\end{equation}
Computing the Poisson bracket of the constraints (\ref{PCDSF}) and
(\ref{SCDSF}),
\begin{equation}\label{PBDSF}
    \{\Phi_D^{(1)}(N),\Phi_D^{(2)}(M)\}=\int_{0}^{\pi}dxM(x)N(x)\not=0,
\end{equation}
we see that the constraints are second class, we have only two
constraints, and if we make a count of degrees of freedom in the
boundary we obtain, zero degrees of freedom, that is the correct
result. Furthermore, the Dirac brackets of the variables are,
\begin{eqnarray}
\{\phi(x,t),\phi(x^{\prime},t)\}^{*}=\{\Pi(x,t),\Pi(x^{\prime},t)\}^{*}=0,\\
\quad \quad \{\phi(x,t),\Pi(x^{\prime},t)\}^{*}=0.
\end{eqnarray}
Then the projection of this brackets to the boundary, results in
the symplectic structure
\begin{eqnarray}
\{\phi(x,t),\phi(x^{\prime},t)\}|_{x=0,\pi}=
\{\Pi(x,t),\Pi(x^{\prime},t)\}|_{x=0,\pi}=0, \label{PB1}\\
 \quad \{\phi(x,t),\Pi(x^{\prime},t)\}|_{x=0,\pi}=\delta(x-x^{\prime})|_{x=0,\pi}=0.
 \label{PB2}
\end{eqnarray}
This structure agree with the results of the exact solution, see
appendix.

\subsection{ Scalar Field with Neumann boundary conditions}
For the Neumann boundary condition (\ref{NSF}), we have the
primary constraint
\begin{equation}\label{PCNSF}
  \partial_{x}\phi(x,t)\approx 0,
\end{equation}
that we can rewrite in the form
\begin{equation}\label{DPCNSF}
 \Phi^{(1)}_N(N)=\int_{0}^{\pi}dxN(x)\partial_{x}\phi(x,t)=0.
\end{equation}
Now the total Hamiltonian is given by
\begin{equation}\label{THSFN}
    H_{T}=H_{c}+\int_{0}^{\pi}dx\lambda(x)\partial\phi(x,t).
\end{equation}
From the evolution of the primary constraint $\Phi^{(1)}_N(N)$ we
obtain the secondary constraint,
\begin{equation}\label{SCNSF}
  \Phi^{(2)}_N(M)=\int_{0}^{\pi}dxM(x)\partial_{x}\Pi(x,t)\approx 0.
\end{equation}
From the stabilization of this constraint results that we don't
have more constraints and the Lagrange multiplier is,
$$\lambda(x)=0.$$
Computing the Poisson bracket between the constraints
$$\{\Phi^{(1)}_N(N),\Phi^{(2)}_N(M)\}=-\int_{0}^{\pi}dxM(x)\partial^{2}_{x}N(x)\not =0,$$
we see that are second class constraints, then we construct the
matrix (\ref{IC}) that in this case have the form,
\begin{eqnarray}
C_{\alpha \beta}(x,x^{\prime})= \left(
\begin{array}{cc}
0 & 1 \\
-1 & 0
\end{array} \right)\partial_{x}\partial_{x^{\prime}}\delta(x-x^{\prime}) \quad {\rm and}\quad
C^{\alpha \beta} = \left(
\begin{array}{cc}
 0 & -1 \\
1 & 0
\end{array} \right)F(x,x^{\prime}),\label{ICNSF}
\end{eqnarray}
where $F(x,x^{\prime})$  is a function with compact support that
satisfies
\begin{eqnarray}
\partial^{2}_{x^{\prime}}F(x,x^{\prime})=-\delta(x-x^{\prime}).\label{eq:tota1}
\end{eqnarray}
The solution to this equation is
\begin{eqnarray}
F(x,x^{\prime})={\Large \sum_{n\geq 1}}\frac{1}{n^{2}\pi}
\sin(nx)\sin(nx^{\prime}).\label{eq:F}
\end{eqnarray}
With the inverse matrix (\ref{ICNSF}), we get the Dirac brackets,
\begin{eqnarray}
\{\phi(x,t),\phi(x^{\prime},t)\}^{*}=\{\Pi(x,t),\Pi(x^{\prime},t)\}^{*}=0,\\
\quad \quad \{\phi(x,t),\Pi(x^{\prime},t)\}^{*}=\Delta(x-x^{\prime})\\
\qquad {\rm with}\qquad
\Delta(x-x^{\prime})=\delta(x-x^{\prime})-\partial_{x}\partial_{x^{\prime}}F(x,x^{\prime}).
\end{eqnarray}
To the function $\Delta(x-x^{\prime})$ we call the Dirac's delta
transverse, because this have the property
$$\partial_{x}\Delta(x-x^{\prime})=\partial_{x^{\prime}}\Delta(x-x^{\prime})=0.$$
This implies
\begin{equation}
\{\partial_{x}\phi(x,t),\Pi(x^{\prime},t)\}^{*}=\{\phi(x,t),
\partial_{x^{\prime}}\Pi(x^{\prime},t)\}^{*}=\{\partial_{x}\phi(x,t),
\partial_{x^{\prime}}\Pi(x^{\prime},t)\}^{*}=0.\nonumber
\end{equation}
So, the symplectic structure that we get for the scalar field with
Neumann boundary conditions is given by,
\begin{eqnarray}
\{\phi(x,t),\phi(x^{\prime},t)\}|_{x=0,\pi}=
\{\Pi(x,t),\Pi(x^{\prime},t)\}|_{x=0,\pi}=0,  \label{PBNSF}\\
\quad
\{\partial_{x}\phi(x,t),\Pi(x^{\prime},t)\}|_{x=0,\pi}=
\partial_{x}\delta(x-x^{\prime})|_{x=0,\pi}=0. \label{PBNSF1}
\end{eqnarray}
Comparing (\ref{PBNSF}-\ref{PBNSF1}) with the result obtained in
the appendix from the exact solution (\ref{ESPBN1}- \ref{ESPBN2}).
We see that both agree completely.

\section{Bosonic string with mixed boundary conditions}
Now we consider the case of the bosonic string in a constant
magnetic field. The action for this string in the conformal gauge
is,
\begin{equation}
S=\frac{1}{2}\int_{-\pi}^{\pi}\int_{t_{1}}^{t_{2}}dxdt[
(\partial_{t} \phi_{i}(x,t))^{2}-(\partial_{x}
\phi_{i}(x,t))^{2}+F_{ij}\partial_{t} \phi_{i}(x,t)\partial_{x}
\phi_{j}(x,t)], \label{ABSCF}
\end{equation}
with $F_{ij}$ an antisymmetric constant matrix, and $i=1,.., N$.
The equations of motion for the system are,
\begin{equation}
(\partial^{2}_{t}- \partial^{2}_{x})\phi_{i}(x,t)=0.
\end{equation}
From the action (\ref{ABSCF}) the canonical Hamiltonian is given
by,
\begin{equation}\label{CHBSCF}
H_{c}=\frac{1}{2}\int_{-\pi}^{\pi}dx[\{\Pi_{i}(x,t)-T_{ij}
\partial_{x}\phi_{i}(x,t)\}^{2} +(\partial_{x}\phi_{i}(x,t))^{2}]
\end{equation}
and we choose Neumann boundary conditions,

\begin{eqnarray}\label{NBM}
 (\partial_{x}\phi_{i} +T_{ij}\partial_{t}
 \phi_{j})(x,t)|_{x=\pm \pi}=0\quad {\rm with}\quad T_{ij}=\frac{F_{ij}}{2}.
\end{eqnarray}
Then, the primary constraints that we have are,
\begin{equation}
\Theta_{i}^{(1)}(x,t)=M_{ij}\partial_{x}\phi_{j}(x,t)+T_{ij}\Pi_{j}(x,t)\approx
0,\quad {\rm with}\quad  M_{ij}=(I-T^{2})_{ji}. \label{eq:pri}
\end{equation}
In the smeared version we have
\begin{equation}
\Theta_{i}^{(1)}[{\bf N}]=\int_{-\pi}^{\pi}dx
N^i(x)\Theta^{(1)}_{i}(x,t). \label{DPCBS}
\end{equation}
The Poisson brackets between these constraints are,
\begin{equation}
\{ \Theta^{(1)}[{\bf N}],\Theta^{(1)}[{\bf M}]\}=0,
\label{eq:numero}
\end{equation}
where we use that the smeared vectors $N^i(x)$ and $M^i(x)$ have
compact support.  From the evolution of these constraints with the
total Hamiltonian, we obtain the secondary constraints,
\begin{equation}
  \Theta^{(2)}[{\bf M}]=\int_{-\pi}^{\pi}dx M^i(x)\partial_{x}\Pi_{i}(x,t)\approx 0.
\end{equation}
The stabilization of these constraints, don't imply new
constraints. So, the complete set of constraints is,
\begin{equation}\label{CBS}
  \chi_\alpha(x,t) =\left(M_{ij}\partial_{x}\phi_{j}(x,t)+T_{ij}\Pi_{j}(x,t)
  ,\ \partial_{x}\Pi_{k}(x,t) \right) \qquad {\rm with} \ \alpha =1,...,2N
\end{equation}
The algebra algebra of these constraints is
\begin{eqnarray}
\{ \chi_\alpha(x,t), \chi_\beta(y,t)\}=C_{\alpha\beta}(x,y),
\end{eqnarray}
where the invertible matrix $C_{\alpha\beta}(x,y)$ is given by,

\begin{eqnarray}
C_{\alpha \beta}(x,y)= \left(
\begin{array}{cc}
0 & M \\
-M & 0
\end{array} \right)\partial_{x}\partial_{y}\delta(x-y).
\end{eqnarray}
For the inverse matrix  $C^{\alpha\beta}(x,y)$ we obtain,
\begin{eqnarray}
\quad C^{\alpha \beta}(x,y) = \left(
\begin{array}{cc}
 0 & -M^{-1} \\
M^{-1} & 0
\end{array} \right)F(x,y), \label{ICBS}
\end{eqnarray}
with $F(x,y)$ defined in (\ref{eq:F}). Using (\ref{ICBS}) we
obtain the Dirac brackets,
\begin{eqnarray}
\{\phi_{i}(x,t),\phi_{j}(y,t)\}^{*}=
-(TM^{-1})_{ij}[\partial_{y}F(x,y)+\partial_{x}F(x,y)],\label{DBPP}\\
\{\Pi_{i}(x,t),\Pi_{j}(y,t)\}^{*}=0,\\
\{\phi_{i}(x,t),\Pi_{j}(y,t)\}^{*}=\delta_{ij}\Delta(x-y).
\end{eqnarray}
Notice that the Dirac bracket (\ref{DBPP}) vanishes in the
boundary, but we have still an extra step in our procedure that
implies to remove from the brackets the contribution that is
different of zero only in the boundary. In order to obtain a Dirac
bracket that is valid only in the bulk we take into account
\rof{F} and comparing with $A(x,y)$ defined in \rof{treta}, we
have
\begin{eqnarray}
\partial_{y}F(x,y)+\partial_{x}F(x,y)=
A(x,y)- \frac{(x+y)}{2\pi}. \label{eq:ser}
\end{eqnarray}
To compare with the exact solution of the original problem we take
the limit of our Dirac brackets to boundary, in this case $x\to
x^{\prime}\to \pm \pi$. So, we have
\begin{eqnarray}
\{\phi_{i}(x=\pm \pi,t),\phi_{j}(x^{\prime}=\pm \pi,t)\}^{*}=\pm
(TM^{-1})_{ij}.
\end{eqnarray}
In consequence using our procedure for the boundary we have the
results
\begin{eqnarray}
\{\phi_{i}(x,t),\phi_{j}(x^{\prime},t)\}|_{x^{\prime}=x=\pm \pi}=\pm (TM^{-1})_{ij}, \\
\{\Pi_{i}(x,t),\Pi_{j}(x^{\prime},t)\}|_{x^{\prime}=x=\pm \pi} =0,\\
\{\partial_{x}\phi_{i}(x,t),\Pi_{j}(x^{\prime},t)\}|_{x^{\prime}=x=\pm
\pi}=\delta_{ij}\partial_{x}
\delta(x-x^{\prime})|_{x^{\prime}=x=\pm \pi} =0.
\end{eqnarray}
with the first derivatives of the deltas null in the boundary.
This result for the symplectic structure in the boundary agrees
with \cite{ho:gnus} and the exact solution
(\ref{EMBC1})-(\ref{EMBC2}) projected in the boundary.

\section{Klein-Gordon equation with mixed boundary conditions}
In this section we analyze the case of the open string ending on a
D-brane in the pp-wave background. This example has been recently
studied in \cite{ho2:gnus}. Where the bosonic action for the open
string in the light cone gauge is
\begin{equation}
S=\frac{1}{2}\int_{-\pi}^{\pi}\int_{t_{1}}^{t_{2}}dxdt[
(\partial_{t} \phi_{i}(x,t))^{2}-(\partial_{x} \phi_{i}(x,t))^{2}-
m^{2}(\phi_{i}(x,t))^{2}+F_{ij}\partial_{t}
\phi_{i}(x,t)\partial_{x} \phi_{j}(x,t)].\label{AOS}
\end{equation}
From this action the canonical Hamiltonian is
\begin{equation}\label{CHOS}
H_{c}=\frac{1}{2}\int_{-\pi}^{\pi}dx[(\Pi_{i}(x,t)-T_{ij}\partial_{x}\phi_{i}(x,t))^{2}
+(\partial_{x}\phi_{i}(x,t))^{2}+m^{2}(\phi_{i}(x,t))^{2}],
\end{equation}
with $T_{ij}$ given in (\ref{NBM}). The primary constraints are
the Neuman boundary conditions,
\begin{equation}\label{PCOS}
\Theta^{(1)}_i(x,t)==M_{ij}\partial_{x}\phi_{j}(x,t)+T_{ij}\Pi_{j}(x,t)\approx
0,
\end{equation}
with $M_{ij}$ given in \rof{pri}. The variation in time of these
constraints produce the secondary constraints
\begin{eqnarray}
 \Theta^{(2)}[{\bf M}]=\int_{-\pi}^{\pi}dxM(x)
 [\partial_{x}\Pi_{i}(x,t)-m^{2}T_{ib}\phi_{b}(x,t))]\approx 0.
\end{eqnarray}
These constraints in the same way that $\Theta^{(1)}[{\bf N}]$
also satisfy
$$\{ \Theta^{(2)}[{\bf N}],\Theta^{(2)}[{\bf M}]\}=0.$$
However, the matrix $C_{\alpha\beta}(x,y)$ given by the Poisson
bracket of all constraints is invertible and have the form
\begin{equation}
C_{\alpha \beta}(x,y)= \left(
\begin{array}{cc}
0 & (M\partial_{y}\partial_{x}-m^{2}T^{2})_{ij} \\
-(M\partial_{y}\partial_{x}-m^{2}T^{2})_{ij} & 0
\end{array} \right).
\end{equation}
Then the inverse matrix $C^{\alpha \beta}(x,y)$ is given
\begin{equation}
C^{\alpha \beta}(x,y) = \left(
\begin{array}{cc}
 0 & -R_{ij}(x,y)  \\
R_{ij}(x,y) & 0
\end{array} \right),
\end{equation}
with $R_{ij}(x,x^{\prime})$ a matrix with compact support, that
satisfies
\begin{eqnarray}
(M\partial^{2}_{x^{\prime}}+m^{2}T^{2})_{ib}R_{bj}(x,x^{\prime})=
-\delta_{ij}\delta(x-x^{\prime}).\label{eq:sta}
\end{eqnarray}
The solution to this boundary problem is,
\begin{eqnarray}
R(x,x^{\prime})={\Large \sum_{n\geq
1}}\frac{\sin(nx)\sin(nx^{\prime})}{\pi [Mn^{2}
-m^{2}T^{2}]}.\label{eq:toti1}
\end{eqnarray}
Then the Dirac brackets are,
\begin{equation}
\begin{split}
\{\phi_{i}(x,t),\phi_{j}(x^{\prime},t)\}^{*}=&
-T_{ia}[\partial_{x^{\prime}}R_{a,j}(x,x^{\prime})
+\partial_{x}R_{a,j}(x,x^{\prime})]\not=0,
 \\
\{\Pi_{i}(x,t),\Pi_{j}(x^{\prime},t)\}^{*}
=&-m^{2}(MT)_{ia}[\partial_{x}R_{aj}(x,x^{\prime})+
\partial_{x^{\prime}}R_{aj}(x,x^{\prime})],\label{eq:dos} \\
\{\phi_{i}(x,t),\Pi_{j}(x^{\prime},t)\}^{*}=&\delta_{ij}(x-x^{\prime})
+m^{2}T^{2}_{ib}R_{bj}(x,x^{\prime})-
\partial_{x}\partial_{x^{\prime}}R_{ib}(x,x^{\prime})M_{bj}.
\end{split}
\end{equation}
Notice that, the massive term is included in all brackets. To
compare with the exact solution, we see that in the term,
\begin{equation}\label{RST}
\begin{split}
\partial_{x}R(x,x^{\prime})+\partial_{x^{\prime}}R_{aj}(x,x^{\prime})=&
{\Large \sum_{n\geq 1}}\frac{n\sin n(x+x^{\prime})}
{\pi [Mn^{2} -m^{2}T^{2}]} \\
=& M^{-1}\left[{\Large \sum_{n\geq 1}}\frac{1}{n\pi}\sin
n(x+x^{\prime}) + {\large \sum_{n\geq 1}}\frac{m^{2}M^{-1}T^{2}
\sin n(x+x^{\prime}) }{n\pi [n^{2} -m^{2}M^{-1}T^{2}]}\right].
\end{split}
\end{equation}
That, using  $A(x,x^{\prime})$ defined in \rof{treta}

\begin{equation}
\partial_{x}R(x,x^{\prime})+\partial_{x^{\prime}}R_{aj}(x,x^{\prime})=
M^{-1}\left[-\frac{(x+x^{\prime})}{2\pi}+A(x,x^{\prime}) + {\Large
\sum_{n\geq 1}}\frac{m^{2}M^{-1}T^{2} \sin n(x+x^{\prime}) }{n\pi
[n^{2} -m^{2}M^{-1}T^{2}]} \right].\nonumber
\end{equation}
Let us remember that before to take the projection to the
boundary, we eliminate all terms that take values only in the
boundary. Then to compare \rof{dos} with the exact solution we
don't take into account the term $A(x,x^{\prime})$. Then taking
the limit $x\to x^\prime \to \pm\pi$ we get

\begin{eqnarray}
[\partial_{x}R(x,x^{\prime})+\partial_{x^{\prime}}R_{aj}(x,x^{\prime})]|_{\pm
\pi} =\pm M^{-1}.
\end{eqnarray}
In this way we arrive to the Dirac brackets projected in the
boundary
\begin{eqnarray}
\{\phi_{i}(x=\pm \pi,t),\phi_{j}(x^{\prime}=\pm \pi,t)\}^{*}&=&\pm (TM^{-1})_{ij},\nonumber\\
\{\Pi_{i}(x=\pm \pi,t),\Pi_{j}(x^{\prime}=\pm \pi,t)\}^{*}&=&\mp  m^{2}T_{ij},\label{ROS}\\
 \{\partial_{x}\phi_{i}(x=\pm \pi,t),\Pi_{j}(x^{\prime}=\pm
 \nonumber
\pi,t)\}^{*}&=&\mp m^{2}(TM^{-1})_{ij}.
\end{eqnarray}
Now, we can compare with the results of \cite{ho2:gnus}, where the
use the exact solution. They get for  $i=1,2$
\begin{eqnarray}
\{\phi_{i}(x,t),\phi_{j}(x^{\prime},t)\}|_{x^{\prime}=x\epsilon \Sigma}&\sim & \pm (TM^{-1})_{ij}, \label{eq:e1} \\
\{\Pi_{i}(x,t),\Pi_{j}(x^{\prime},t)\}|_{x^{\prime}=x\epsilon \Sigma}&\sim & \pm m^{2}T_{ij},\label{eq:e2}\\
\{\phi_{i}(x,t),\Pi_{j}(x^{\prime},t)\}&=&\delta_{ij}\delta(x-x^{\prime})\label{eq:e3}
\end{eqnarray}
with $\partial_{x}\delta(x-x^{\prime})|_{x\epsilon \Sigma}=0.$
Here we see that our results agree in \rof{e1} and \rof{e2} but
not for \rof{e3}. However is easy to see that the result \rof{e3}
is inconsistent with the boundary condition. Taking the partial
derivative with respect to $x$ in \rof{e3}, we get

$$
\{\partial_{x}\phi_{i}(x,t),\Pi_{j}(x^{\prime},t)\}|_{x\epsilon
\Sigma}=\delta_{ij}\partial_{x}\delta(x-x^{\prime})|_{x\epsilon
\Sigma}=0.$$ On the other hand using the boundary condition
(\ref{PCOS}) in \rof{e2} we get
$$
\{\partial_{x}\phi^{i}(x,t),\Pi^{j}(x^{\prime},t)\}|_{x\epsilon
\Sigma}\sim \pm m^{2}(M^{-1}T^{2})^{ij}.$$ Then, the results
\rof{e2} and \rof{e3} are inconsistent, whereas (\ref{ROS}) is
fully consistent with the boundary conditions.

\section{Conclusions}
In the present paper was developed a procedure to compute the
symplectic structure for a field theory with boundary. We consider
that the boundary conditions can be interpreted as Dirac's
constraints and we construct a procedure that following the
ordinary steps of the Dirac's method, produce the symplectic
structure on the boundary, for several examples. In our procedure,
we have the problem that does not exist a definition for a smeared
Dirac bracket so to compute this bracket we consider the
prescription that the Dirac bracket in the interior of $\Sigma$ is
given by the ordinary Dirac bracket valued on $\Sigma$ minus the
contributions that are no null only in the boundary
$\partial\Sigma$. Using this prescription we are capable to
compute the symplectic structure in several examples always
obtaining fully consistent results. In a future we will try to
extend our procedure to the case of General Relativity and compare
our results with the recently published papers \cite{Anco1:gnus,
Anco2:gnus, Pons:gnus}

\acknowledgments The authors acknowledge partial support from DGAPA-UNAM grant
IN117000.

\appendix
\section{Appendix}\label{sec:A}

\setcounter{section}{1}

In this appendix, we show the results obtained by using the exact
solutions.

\subsection{Scalar field}
The general solution of the equation of motion for the scalar
field (\ref{EMSF}) with Dirichlet boundary conditions (\ref{DSF})
is
\begin{eqnarray}
\phi(x,t)={\Large \sum_{n\geq 1}
}q_{n}(t)\sqrt{\frac{2}{\pi}}\sin(nx),\label{eq:ajo}
\end{eqnarray}
where $q_n(t)$ satisfies
$$\ddot q_{n}(t)=-n^{2} q_{n}(t).$$
The inverse relation to \rof{ajo} is given by
\begin{eqnarray}
 q_{n}(t)=\sqrt{\frac{2}{\pi}}\int_{0}^{l}dx \phi(x,t)\sin(nx).\label{eq:cebolla}
\end{eqnarray}
Using \rof{cebolla} in \rof{ajo} we found
$$\phi(x,t)=\int_{0}^{\pi}dx^{\prime} \phi(x^{\prime},t)\frac{2}{\pi}{\Large \sum_{n\geq 1 }}
\sin(nx^{\prime})\sin(nx)=\int_{0}^{l}dx^{\prime}
\phi(x^{\prime},t)\delta(x-x^{\prime}),$$ that means that the
Dirac delta is,
$$\delta(x-x^{\prime})=\frac{2}{\pi}{\Large \sum_{n\geq 1}}\sin(nx^{\prime})\sin(nx).$$
On the other hand, the canonical momentum is,
\begin{eqnarray}
\Pi(x,t)=\frac{\partial L}{\partial\dot \phi(x,t)}=\dot
\phi(x,t)={\Large \sum_{n\geq 1}} \dot
q_{n}(t)\sqrt{\frac{2}{\pi}}\sin(nx),\label{eq:chile}
\end{eqnarray}
form this follows the notation $\dot q_{n}(t)=p_{n}(t)$, and then
we have
\begin{eqnarray}
\Pi(x,t)=\sqrt{\frac{2}{\pi}}{\Large \sum_{n\geq 1}}
p_{n}(t)\sin(nx)
\end{eqnarray}
with inverse given by
\begin{eqnarray}
\qquad p_{n}(t)=\sqrt{\frac{2}{\pi}}\int_{0}^{\pi}dx
\Pi(x,t)\sin(nx).
\end{eqnarray}
Now, considering that the Poisson brackets between the normal
modes are,
$$\{q_n(t),q_m(t)\}=\{p_n(t),p_m(t)\}=0
\qquad {\rm and}\qquad \{q_n(t),p_m(t)\}=\delta_{nm}.$$ In this way,
the Poisson brackets that follows from the exact solution are,
\begin{equation}\label{ESSF}
\begin{split}
  \{\phi(x,t),\Pi(x^{\prime},t)\}&=\frac{2}{\pi} {\Large \sum_{n\geq
1}}\sin(nx^{\prime})\sin(nx)=\delta(x-x^{\prime}) \\
\{\phi(x,t),\phi(x^{\prime},t)\}&=\{\Pi(x,t),\Pi(x^{\prime},t)\}=0.
\end{split}
\end{equation}
The projection of this result to the boundary, gives exactly the
same result, that we get using our procedure, see (\ref{PB1}) and
(\ref{PB2}).

For the Neumann boundary conditions, using the exact solution we
get

\begin{eqnarray}
\{\phi(x,t),\Pi(x^{\prime},t)\}&=&\frac{1}{\pi}+\frac{2}{\pi}
{\Large \sum_{n\geq 1}}{\rm cos}(nx^{\prime})
{\rm cos}(nx)=\delta(x-x^{\prime}) \label{ESPBN1}\\
 \{\partial_{x}\phi(x,t),\Pi(x^{\prime},t)\}|_{x=\pi,0}&=&0=
 \partial_{x}\delta(x-x^{\prime})|_{x=\pi,0}. \label{ESPBN2}
\end{eqnarray}
If we compare this result with our method of computation
(\ref{PBNSF}) and (\ref{PBNSF1}) we see that both result agree.
So, we see that in the case of the scalar field the projection of
the Dirac brackets obtained from our procedure agrees completely
with the result obtained using the exact solution.

\subsection{Mixed Conditions}
The general solution for the equations of motion in this case is
of the form,

$$\phi_{i}(x,t)=M_{i}+B_{i}t+T_{ij}B_{j}x +{\Large \sum_{n\geq 1} }\frac{q_{ni}(t)}
{\sqrt{\pi}}{\rm cos}(nx)-\frac{T_{ij}p_{nj}(t)}{n\sqrt{\pi}}\sin(nx).$$
With $\ddot q_{i}(t)=-q_{i}(t)$ and $\dot q_{i}=p_{i}(t)$.
Furthermore, from the definition of the canonical momenta we have
$$\Pi_{i}(x,t)=M_{ij}[B_{j}+{\Large \sum_{n\geq 1} }\frac{p_{nj}(t)}
{\sqrt{\pi}}{\rm cos}(nx)].$$ Then, follows that the Fourier
coefficients are,

\begin{eqnarray}
q_{in}(t)&=&\frac{1}{\sqrt{\pi}}\int_{-\pi}^{\pi}dx \phi(x,t){\rm cos}(nx),\\
 p_{in}(t)&=&\frac{1}{\sqrt{\pi}}\int_{-\pi}^{\pi}dx M^{-1}_{ij}\Pi_{j}(x,t){\rm cos}(nx),\\
 B_{i}&=&\frac{2}{\pi}\int_{-\pi}^{\pi}dx M^{-1}_{ij}\Pi_{j}(x,t), \\
A_{i}&=&\frac{2}{\pi}\int_{-\pi}^{\pi}dx[ \phi_{i}(x,t)-
M^{-1}_{ij}\Pi_{j}(x,t)].
\end{eqnarray}
For the Poisson brackets we get,
$$\{q_{in}(t),p_{jm}(t)\}=M^{-1}_{ij}\delta_{nm},\qquad \{A_{i},B_{j}\}=
\frac{M^{-1}_{ij}}{2\pi},$$ and the brackets vanish for the other
cases. Using, the exact solution we obtain

\begin{equation}\label{EMBC1}
\{\phi_{i}(x,t),\Pi_{j}(x^{\prime},t)\}=\delta_{ij}\left[\frac{1}{2\pi}+
\frac{1}{\pi}{\Large \sum_{n\geq 1}}{\rm cos}(nx^{\prime}){\rm
cos}(nx)\right]=\delta_{ij}\delta(x-x^{\prime}), \end{equation}and

\begin{equation}\label{EMBC2}
\{\phi_{i}(x,t),\phi_{j}(x^{\prime},t)\}=(M^{-1}T)_{ij}\left[\frac{(x+x^{\prime})}
{2\pi}+{\Large \sum_{n\geq 1}}\frac{\sin
n(x+x^{\prime})}{n\pi}\right], \end{equation} In particular on the
boundary
$$\{\phi_{i}(\pm \pi,t),\phi_{j}(\pm \pi,t)\}=\pm(M^{-1}T)_{ij}.$$
and we see that the fields do not commute on the boundary, whereas
the canonical momenta are commutative.\\

On the other hand, from the consistency with the Poisson brackets
we see that
\begin{eqnarray}
A(x,x^{\prime})= \left[\frac{(x+x^{\prime})}{2\pi}+{\Large
\sum_{n\geq 1}}\frac{\sin n(x+x^{\prime})}{n\pi}\right],
\label{eq:treta}
\end{eqnarray}
must be vanish in all interval  $(-\pi,\pi)$, except a zero
measure set, in the boundary satisfies
$$A(\pm \pi,\pm \pi)=\pm 1.$$
In fact we see that $A$ is a distribution given by
$$\partial_{x} A(x,x^{\prime})=\partial_{x^{\prime}}
A(x,x^{\prime})=\delta_{a}(x-x^{\prime})-\delta_{b}(x-x^{\prime}).$$
With

$$\delta_{a}(x-x^{\prime})=\frac{1}{2\pi}+\frac{1}{\pi}{\Large
\sum_{n\geq 1}}{\rm cos}(nx^{\prime}){\rm cos}(nx),$$ and

$$\delta_{b}(x-x^{\prime})=\frac{1}{\pi}{\Large \sum_{n\geq 1}}
\sin(nx^{\prime})\sin(nx).$$

\end{document}